\documentclass{aa}  
\usepackage{graphicx}
\usepackage{txfonts}
\usepackage{mathrsfs}
\begin{document}

\title{Cosmological parameters from strong gravitational lensing and
  stellar~dynamics in elliptical galaxies}  

\author{C. Grillo \inst{1,2}, M. Lombardi \inst{1,2}, and G. Bertin \inst{2}}

\offprints{C. Grillo}

\institute{European Southern Observatory, Karl-Schwarzschild-Str. 2,
  D-85748, Garching bei M\"unchen, Germany\\
  \email{cgrillo@eso.org}
  \and
  Universit\`a degli Studi di Milano, Department of Physics,
  via Celoria 16, I-20133 Milan, Italy\\
}

\authorrunning{C.~Grillo et al.}

\titlerunning{Cosmological parameters from strong gravitational
  lensing and stellar dynamics in elliptical galaxies}

\date{Received X X, X; accepted Y Y, Y}

\abstract
{Observations of the cosmic microwave background, light element
  abundances, large-scale distribution of galaxies, and distant
  supernovae are the primary tools for determining the cosmological
  parameters that define the global structure of the Universe.} 
{Here we illustrate how the combination of observations related to
  strong gravitational lensing and stellar dynamics in elliptical
  galaxies offers a simple and promising way to measure the cosmological
  matter and dark-energy density parameters.}
{A gravitational lensing estimate of the mass enclosed inside the
  Einstein circle can be obtained by measuring the Einstein
  angle, once the critical density of the system is known. A
  model-dependent dynamical estimate of this mass can also be obtained
  by measuring the central velocity dispersion of the stellar
  component. By assuming the well-tested homologous $1/r^{2}$
  (isothermal) profile for the total (luminous+dark) density
  distribution in elliptical galaxies acting as lenses, these two mass
  measurements can be properly compared. Thus, a relation between the
  Einstein angle and the central stellar velocity dispersion is
  derived, and the cosmological matter and the dark-energy
  density parameters can be estimated from this.}
{We determined the accuracy of the cosmological parameter estimates by
  means of simulations that include realistic measurement
  uncertainties on the relevant quantities. Interestingly, the 
  expected constraints on the cosmological parameter plane are
  complementary to those coming from other observational
  techniques. Then, we applied the method to the recent data sets of the
  \emph{Sloan Lens ACS} (SLACS) and the \emph{Lenses Structure and
  Dynamics} (LSD) Surveys, and showed that the concordance value between
  0.7 and 0.8 for the dark-energy density parameter is included in our
  99\% confidence regions.}
{The small number of lenses available to date prevents us from precisely determining the cosmological parameters, but it still proves
  the feasibility of the method. When applied to samples made of
  hundreds of lenses that are expected to become available from
  forthcoming deep and wide surveys, this technique will be an
  important alternative tool for measuring the geometry of the Universe.}

\keywords{cosmology: theory -- cosmology: observations -- galaxies:
	distances and redshifts -- galaxies: elliptical and
	lenticular, cD -- gravitational lensing -- galaxies:
	kinematics and dynamics}

\maketitle
%

\section{Introduction}

The Universe appears to be dominated by dark-energy and dark
matter. Although the physical nature of these dark components is still
unknown, the standard cosmological $\Lambda$\emph{CDM} model with only
a few parameters fits most of the current data well: precision
measurements of the anisotropies in the cosmic microwave background
(\emph{CMB}; Bennett et al. \cite{ben}; Spergel et al. \cite{spe1},
\cite{spe2}), the observed abundances of light elements (Burles et
al. \cite{bur}; Cyburt et al. \cite{cyb}), the large-scale
distribution of galaxies (\emph{LSS}; Tegmark et al. \cite{teg}; Cole
et al. \cite{col}), and the luminosity-distance relationship for
distant type Ia supernovae (\emph{SNIa}; Riess et al. \cite{rie1},
\cite{rie2}; Perlmutter et al. \cite{per}). In this standard model,
the Universe is homogeneous and isotropic on its largest scales, and
its geometry appears to be flat ($\Omega = \Omega_{m} +
\Omega_{\Lambda} \approx 1$); the total mass-energy density is mainly
in the form of dark-energy ($\Omega_{\Lambda} \approx 0.7$) and matter
($\Omega_{m} \approx 0.3$), ordinary and dark. These values of the
cosmological parameters imply a fairly recent transition from a
decelerating to an accelerating universal expansion. The 
\emph{Hubble Space Telescope} (\emph{HST}) \emph{Key Project} has
measured the current expansion rate, the Hubble parameter $H_{0}= (72
\pm 8)\,\mathrm{km\,s^{-1}\,Mpc^{-1}}$ (Freedman et
al. \cite{fre}). The estimates of different cosmological parameters
from a single observational method are often correlated (hence
``degenerate'') and exhibit significant uncertainties. For instance,
from the \emph{WMAP} three year data alone, without a prior on the
flatness of the Universe, the best-fit model is characterized by
$\Omega_{m} = 0.42$, $\Omega_{\Lambda} = 0.63$,
$H_{0}=55\,\mathrm{km\,s^{-1}\,Mpc^{-1}}$ (Spergel et
al. \cite{spe2}), values that are quite different from the concordance
values reported above. This suggests that precise measurements of the
cosmological parameters can only be obtained by using complementary
techniques. In fact, considerable efforts are still being made in
order to secure accurate measurements of these parameters (in
particular, see the scientific goals of the forthcoming \emph{PLANCK}
and \emph{SNAP} missions).

The deflection of light due to gravitational lensing is sensitive to
the total matter density of the structures in the Universe,
independently of the nature or dynamical state of the deflecting
mass. Therefore, strong and weak gravitational lensing provide
valuable tools for measuring the distribution of mass. In particular,
cosmic shear estimates of the amplitude of the weak lensing
distorsions of distant sources over a wide range of angular scales
(Bartelmann \& Schneider \cite{bar2}) are a very encouraging way to
study the large-scale structure of the Universe, and therefore to
probe the parameters that define the relevant cosmological model
(Refregier \cite{ref}). Further constraints on the geometry of the
Universe can also be provided by the abundance of lenses or arcs
observed in lens surveys (see Bartelmann \& Weiss \cite{bar1} for a numerical approach; Mitchell et al. \cite{mit} for observational results from the CLASS Survey described by Myers et al. \cite{mye} and Browne et al. \cite{bro}) and by strong (Link \& Pierce \cite{lin}; Soucail et
al. \cite{sou}) or weak (Lombardi \& Bertin \cite{lom}; Jain \& Taylor
\cite{jai}) lensing mass reconstructions in clusters of
galaxies. Isolated strong lens galaxies offer another possibility to 
measure the cosmological parameters (Kochanek \cite{koc3},
\cite{koc4}; Myungshin et al. \cite{myu}), including the Hubble
parameter (Refsdal \cite{refs}; Koopmans et al. \cite{koo4};
M\"ortsell \& Sunesson \cite{mor}).

The various techniques that have been proposed are limited by several
assumptions. For example, a measurement of the matter density
parameter from cosmic shear is degenerate with that of the 
normalization of the amplitude of the power spectrum of matter 
perturbations ($\sigma_{8}$); moreover, an extremely large number of
high-quality galaxy images and some modeling on the growth of the
structure in the Universe are required. For techniques based on
gravitational lens statistics, the luminosity function, the relation
between luminosity and velocity dispersion, and the density profile of
the lens galaxies play an important role. The use of arcs statistics
awaits more realistic simulations of clusters and observations, since
different studies have led to contrasting results (e.g., Bartelmann et
al. \cite{bar3}; Dalal et al. \cite{dal}). The estimates of the
cosmological parameters from cluster mass reconstructions have also
two distinct limitations: the presence of possible substructure in the
region of multiple image formation (for strong lensing) and the need
for a high density of background galaxies (for weak lensing).

In this paper we propose a new technique that, starting from strong
gravitational lensing and stellar dynamics observations in elliptical
galaxies, is able to probe the geometry of the Universe in a different
and effective way. The paper is organized as follows. In Sect.~2, we
describe our method to estimate the matter and dark-energy density
parameters. Then, in Sect.~3, we determine through simulations the
precision attainable in these measurements of the cosmological
parameters. Good estimators for the quantities relevant to the problem
are identified in Sect.~4. In Sect.~5, we apply our technique to the
data collected and just published of two surveys of lens galaxies for
which stellar dynamical measurements are available. Finally, in
Sect.~6, we summarize and discuss the results obtained.

\section{The method}

It is known that in an axisymmetric lens multiple images can only form
in the vicinity of the so-called Einstein ring, at an angle
$\theta_{\mathrm{E}}$ from the center of the lens (see Schneider et
al. \cite{schneider}). From the theory of gravitational lensing, the
``mass'' $\mathscr{M}_{\mathrm{grl}}$ enclosed within the disk defined
by the Einstein ring is directly related to the geometry of the
configuration, through the definition of the critical density
($\mathscr{M}_{\mathrm{grl}} =
\Sigma_{\mathrm{cr}}\pi\theta^{2}_{\mathrm{E}}$). This ``mass''
$\mathscr{M}_{\mathrm{grl}}$ is connected to the intrinsic mass
$M_{\mathrm{grl}}$ of the lens by the distance to the lens (by
converting the Einstein angle into an Einstein radius).

A dynamical estimate $\mathscr{M}_{\mathrm{dyn}}$ of the mass can also
be given by measuring the quantity
$\sigma_{0}^{2}\theta_{\mathrm{E}}$, where $\sigma_{0}$ is the central
velocity dispersion of the stellar component, usually referred to the
disk of radius $R_{\mathrm{e}}/8$ ($R_{\mathrm{e}}$ being the standard
optical effective radius); the dynamical mass is then obtained by
multiplication by a suitable factor ($\mathscr{M}_{\mathrm{dyn}} =
\alpha\,\sigma_{0}^{2}\theta_{\mathrm{E}}$) that is model-dependent
(the lens usually includes a significant dark matter
component). Again, in order to relate $\mathscr{M}_{\mathrm{dyn}}$ to
the intrinsic mass $M_{\mathrm{dyn}}$ we should convert
$\theta_{\mathrm{E}}$ into a radius.

With no need to refer to intrinsic masses (and thus with no need to
know the exact distance to the lens, which would bring in a knowledge
of the Hubble constant $H_{0}$), we thus see that, if we identify
$\mathscr{M}_{\mathrm{grl}} = \mathscr{M}_{\mathrm{dyn}}$, the
combination of a measurement of $\theta_{\mathrm{E}}$ and of
$\sigma_{0}$ should be uniquely related, in the standard cosmological
model, to a function of the redshifts $z_{l}$ and $z_{s}$ (of the lens
and of the source, respectively) and of the cosmological parameters
$\Omega_{m}$ and $\Omega_{\Lambda}$. Note that, in this context,
$z_{l}$ and $z_{s}$ can be considered to be measured with negligible
errors.

Several studies, based on various dynamical tracers (stars, globular clusters, planetary nebulae, X-ray halos, HI disks and rings), have established that bright elliptical galaxies of the local universe, as a rule, exhibit approximately flat circular velocity curves (e.g., Gerhard et al. \cite{ger}; Peng et al. \cite{pen}), thus suggesting that the structure of these systems should be considered as approximately homologous, with the total density distribution (luminous+dark) close to that of a singular isothermal sphere (SIS; $\rho \propto 1/r^2$). These detailed dynamical studies refer to nearby galaxies and thus the result obtained does not depend on the values of the cosmological parameters. Of course, this is a zeroth-order description, and different galaxies may exhibit different deviations from this ``universal" total density profile.

A number of investigations of galaxies at cosmologically significant distances address observed properties that result from the combined effects of evolution and of the geometry of the universe. The interpretation of these data can thus be obtained in different ways. For example, in the study of the Fundamental Plane out to $z \approx 1$ (e.g., see Treu et al. \cite{tre5} and other parallel investigations, many of which quoted there), one may assume approximate structural homology and a given cosmological model and use the data on the observed change in the Fundamental Plane to derive information on the evolution properties of the observed stellar populations.

Here we recall that, under the assumption of the concordance cosmological model ($\Omega_m = 0.3~,~\Omega_{\Lambda} = 0.7~,~ H_0 = 70~\mathrm{km~s^{-1}~Mpc^{-1}}$), analyses of strong gravitational lensing alone (e.g., Rusin e al. \cite{rus}) or combined measurements of stellar dynamics and gravitational lensing (e.g., Treu \& Koopmans \cite{tre2}; Koopmans et al. \cite{koo3}) have confirmed the persistence of structural homology, i.e., that little evolution appears to take place in the observed total density profile of bright ellipticals, in the sense that they are found to be characterized by approximately SIS density profiles also at cosmologically significant distances. In Sect.~5 we will show that this conclusion is robust with respect to the choice of the adopted cosmological parameters.

These findings have encouraged us to explore the consequences of considering the combined measurements of stellar dynamics and gravitational lensing on distant ellipticals, starting from the simplifying assumption that homology is indeed strictly followed by these systems; in particular, we wish to explore whether this assumption, applied to the interpretation of the data, may lead to interesting constraints on the values of the parameters that define the geometry of the universe. In other words, we wish to check the consequences on the cosmological parameters of assuming from the very beginning that (1) the espression for $\mathscr{M}_{\mathrm{grl}}$ is basically that associated with an SIS, and (2) no significant variation on the ``virial coefficient" $\alpha$ is present from galaxy to galaxy.

In practice, we proceed as follows. We note that for an SIS
\begin{equation}
\theta_{\mathrm{E}}=4\pi\,\bigg(\frac{\sigma_{\mathrm{SIS}}}{c}\bigg)^{2}\frac{D_{ls}}{D_{os}}\,,
\label{eq:1}
\end{equation} 
where $\sigma_{\mathrm{SIS}}$ is the lens ``velocity dispersion'', $c$
is the light speed, $D_{ls}$ and $D_{os}$ are the lens-source and the
observer-source angular diameter distances, respectively. We then
bypass the issues raised by dynamical modeling by recalling that
$\sigma_{0}$ turns out to be a good estimate of
$\sigma_{\mathrm{SIS}}$. This latter point, exploited by Kochanek
(\cite{koc1}, \cite{koc2}), was confirmed by Treu et al. (\cite{tre1}), and is now
checked by us independently, by a test described separately in
Sect. 4.2 on a sample of eight well-studied nearby
ellipticals. Therefore we consider the quantity  
\begin{equation}
\frac{c^{2}}{4\pi}\,\frac{\theta_{\mathrm{E}}}{\sigma_{0}^{2}}=\frac{D_{ls}}{D_{os}}=r(z_{l},z_{s};\Omega_{m},\Omega_{\Lambda})\,,
\label{eq:2}
\end{equation} 
as the observable that will be used to produce, by studying a
statistically significant sample of lenses, a measurement of
$\Omega_{m}$ and $\Omega_{\Lambda}$.

Given the weak dependence of the relevant coefficients, such as
$\alpha$, on the detailed relative distributions of dark and luminous
matter in galaxies, the method is expected to be robust. In
particular, the entire argument could be easily generalized to
non-axisymmetric lenses by referring to the properties of the
so-called singular isothermal ellipsoid (see Kormann et
al. \cite{kor}).

\begin{figure}
  \centering
  \includegraphics[width=0.50\textwidth]{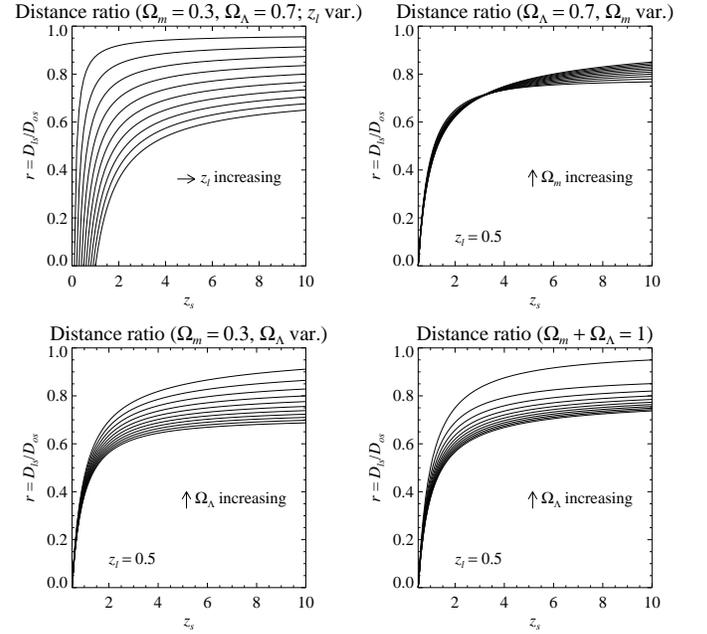}
  \caption{Dependence of the angular diameter distance ratio
  $r=D_{ls}/D_{os}$ on the lens redshift (\emph{top left}), matter
  density parameter (\emph{top right}), dark-energy density parameter
  (\emph{bottom left}), and matter and dark-energy density parameters
  in a flat cosmological model (\emph{bottom right}). In the first
  panel, $(\Omega_{m},\Omega_{\Lambda}) = (0.3,0.7)$ are fixed and
  $z_{l}$ is varied from 0.1 to 1, with a regular step of 0.1. The
  arrow shows the increasing direction of $z_{l}$. In the second
  panel, $\Omega_{\Lambda} = 0.7$ and $z_{l}=0.5$ are fixed and
  $\Omega_{m}$ is varied from 0 to 1, with a regular step of 0.1. The
  arrow shows the increasing direction of $\Omega_{m}$. In the third 
  panel, $\Omega_{m} = 0.3$ and $z_{l}=0.5$ are fixed and
  $\Omega_{\Lambda}$ is varied from 0 to 1, with a regular step of
  0.1. The arrow shows the increasing direction of
  $\Omega_{\Lambda}$. In the fourth panel, $\Omega_{m} + 
  \Omega_{\Lambda} = 1$ and $z_{l}=0.5$ are fixed and
  $\Omega_{\Lambda}$ is varied from 0 to 1, with a regular step of
  0.1. The arrow shows the increasing direction of
  $\Omega_{\Lambda}$.}
  \label{teo}
\end{figure}

Figure \ref{teo} illustrates the distance ratio $r$ of
Eq. (\ref{eq:2}) (which does not have a simple analytic form) versus
the source redshift, as a function of the lens redshift and of the
matter and dark-energy density parameters. In general, different
cosmological models give values of $r$ which differ more clearly at
higher values of the source redshift (see the last three panels); in
addition, the quantity $r$ is more sensitive to small variations of
$\Omega_{\Lambda}$ than of $\Omega_{m}$ (compare the second and third
panels). As a consequence, we naturally expect this method to be
optimally efficient in measuring $\Omega_{\Lambda}$, provided we have
at our disposal a sample of lenses at sufficiently high redshifts.

\section{Simulated measurements of the cosmological parameters}

\begin{figure*}
  \centering
  \includegraphics[width=\textwidth]{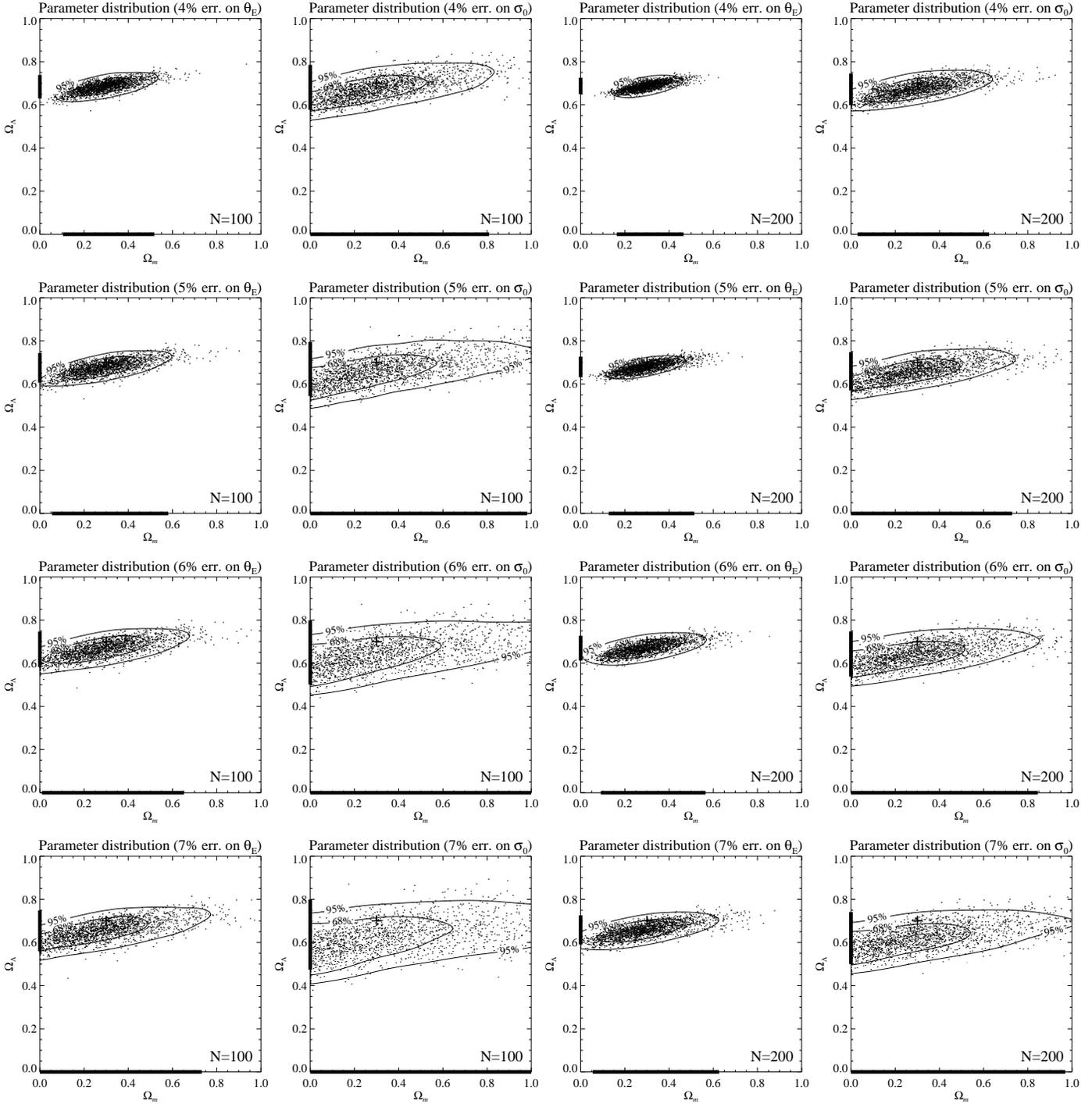}
  \caption{Estimates of the cosmological parameters. Simulation of 2000
  measurements composed of 100 (\emph{on the left}) and 200 (\emph{on
  the right}) lenses each, with different uncertainties (increasing
  from the top to the bottom) on the Einstein angle (\emph{first and
  third column}) and on the central velocity dispersion (\emph{second
  and fourth column}). A nominal $0\%$ uncertainty is assigned to the
  quantity not mentioned in the panels. Thick bars on the co-ordinate
  axes and contour levels on the planes represent, respectively, the
  95\% confidence intervals and the 68\% and 95\% confidence regions
  for the cosmological parameters. A cross shows the position of the
  true parameters: $(\Omega_{m},\Omega_{\Lambda}) = (0.3,0.7)$.}
  \label{pa1}
\end{figure*}

\begin{figure*}
  \centering
  \includegraphics[width=\textwidth]{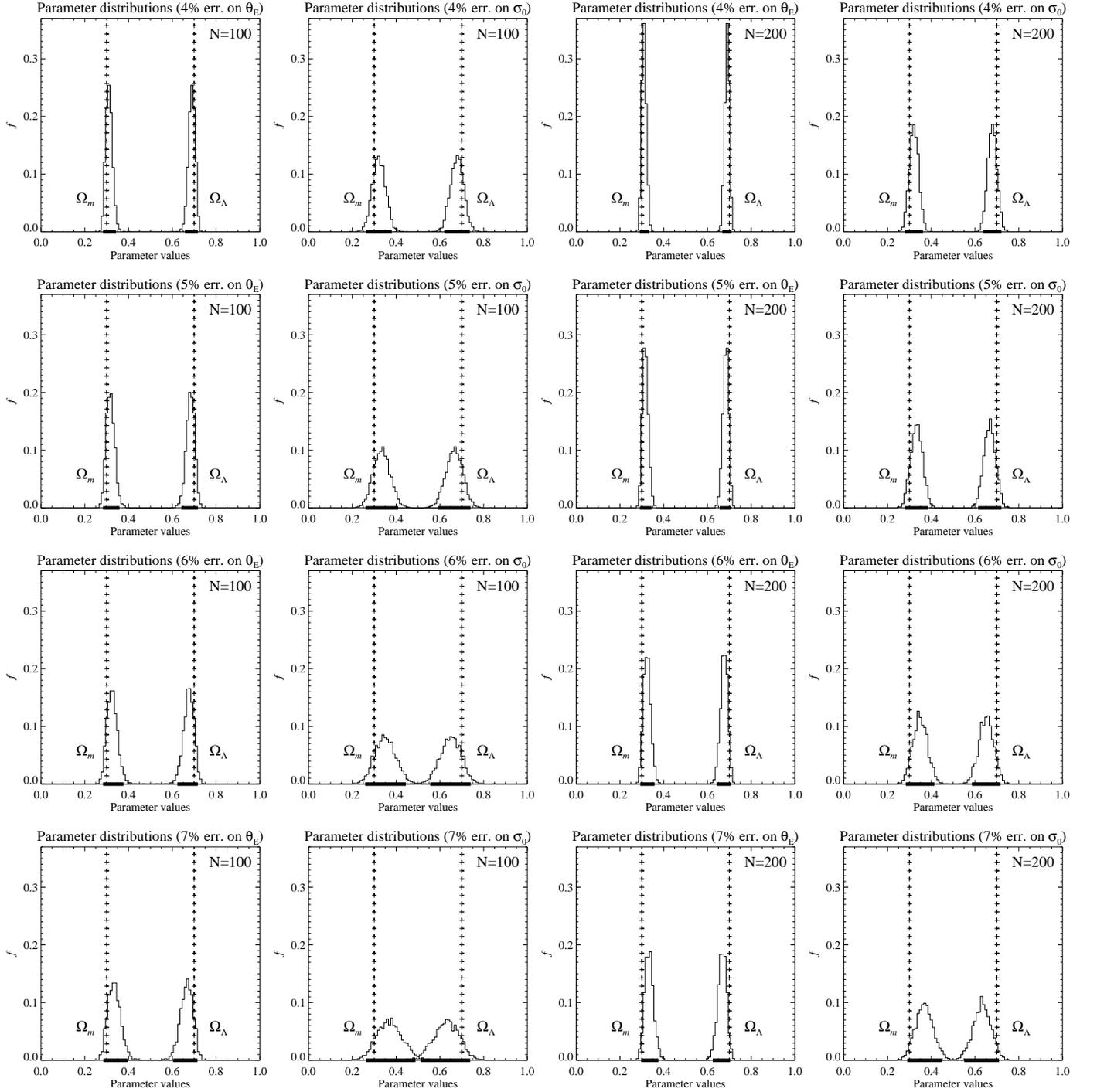}
  \caption{Estimates of the cosmological parameters in a flat model
  ($\Omega_{m}+\Omega_{\Lambda}=1$). Simulation of 2000 
  measurements composed of 100 (\emph{on the left}) and 200 (\emph{on
  the right}) lenses each, with different uncertainties (increasing
  from the top to the bottom) on the Einstein angle (\emph{first and
  third column}) and on the central velocity dispersion (\emph{second
  and fourth column}). A nominal $0\%$ uncertainty is assigned to the
  quantity not mentioned in the panels. Thick bars on the abscissa
  axes represent the 95\% confidence intervals for the cosmological
  parameters. Crosses show the position of the true parameters:
  $(\Omega_{m},\Omega_{\Lambda}) = (0.3,0.7)$.} 
  \label{pa2}
\end{figure*}

\begin{figure}
  \centering
  \includegraphics[width=0.48\textwidth]{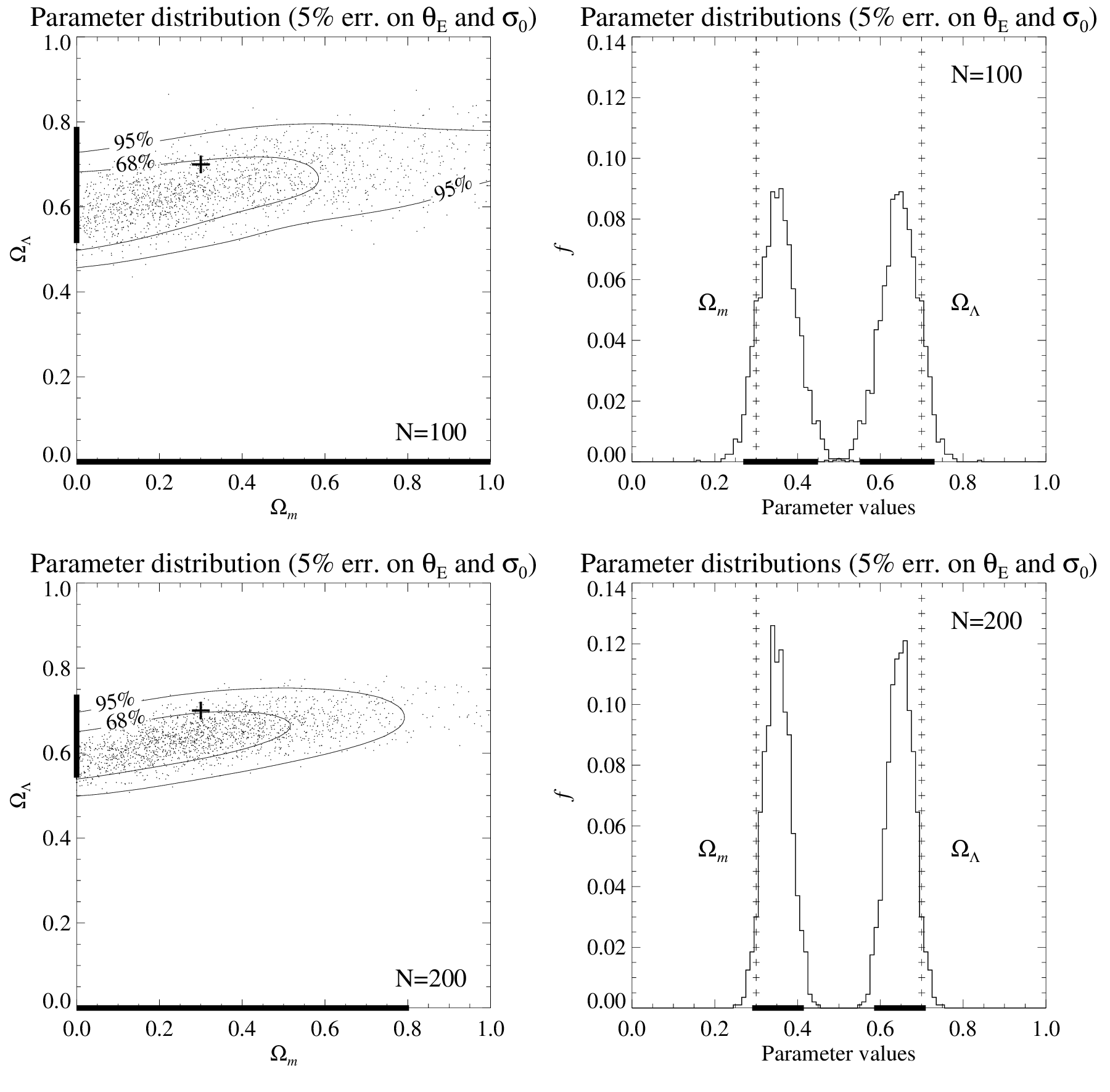}
  \caption{Estimates of the cosmological parameters assuming 5\%
  uncertainty on both the Einstein angle and the central velocity
  dispersion. Simulation of 2000 measurements composed of 100
  (\emph{on the top}) and 200 (\emph{on the bottom}) lenses each, in
  general (\emph{on the left}) and in flat (\emph{on the right})
  cosmological models. Thick bars on the co-ordinate axes and contour
  levels on the planes represent, respectively, the 95\% confidence
  intervals and the 68\% and 95\% confidence regions for the
  cosmological parameters. Crosses show the position of the true
  parameters: $(\Omega_{m},\Omega_{\Lambda}) = (0.3,0.7)$.} 
  \label{pa3}%
\end{figure}

\begin{table*}
\caption{Intervals at 95\% confidence level for the matter and the
  dark-energy density parameters.} 
\label{table:1}
\centering
\begin{tabular}{c c c c c c c c c c}
\hline\hline & & N=100 & & & & & N=200 & & \\ \hline
Err. $\theta_{\mathrm{E}}$ & $4\%$ & $5\%$ & $6\%$ & $7\%$ &
Err. $\theta_{\mathrm{E}}$ & $4\%$ & $5\%$ & $6\%$ & $7\%$ \\ \hline
$\Omega_{m}$ & $\left[0.10,0.52\right]$ & $\left[0.06,0.58\right]$ &
$\left[0.01,0.65\right]$ & $\left[0.00,0.73\right]$ & $\Omega_{m}$ &
$\left[0.16,0.47\right]$ & $\left[0.13,0.51\right]$ &
$\left[0.09,0.57\right]$ & $\left[0.05,0.63\right]$ \\ 
$\Omega_{\Lambda}$ & $\left[0.63,0.74\right]$ &
$\left[0.61,0.74\right]$ & $\left[0.58,0.75\right]$ &
$\left[0.56,0.75\right]$ & $\Omega_{\Lambda}$ &
$\left[0.65,0.73\right]$ & $\left[0.63,0.73\right]$ &
$\left[0.61,0.73\right]$ & $\left[0.59,0.73\right]$ \\ \hline
Err. $\sigma_{0}$ & $4\%$ & $5\%$ & $6\%$ & $7\%$ &
Err. $\sigma_{0}$ & $4\%$ & $5\%$ & $6\%$ & $7\%$ \\ \hline
$\Omega_{m}$ & $\left[0.00,0.81\right]$ & $\left[0.00,0.98\right]$ &
$\left[0.00,1.19\right]$ & $\left[0.00,1.41\right]$ & $\Omega_{m}$ &
$\left[0.03,0.63\right]$ & $\left[0.00,0.73\right]$ &
$\left[0.00,0.85\right]$ & $\left[0.00,0.97\right]$ \\
$\Omega_{\Lambda}$ & $\left[0.58,0.78\right]$ &
$\left[0.54,0.80\right]$ & $\left[0.50,0.80\right]$ &
$\left[0.47,0.80\right]$ & $\Omega_{\Lambda}$ &
$\left[0.60,0.75\right]$ & $\left[0.57,0.75\right]$ &
$\left[0.54,0.75\right]$ & $\left[0.50,0.74\right]$ \\ \hline
\end{tabular}
\begin{list}{}{}
\item[Notes --]These intervals are obtained by
  excluding from the 2000 $\chi^{2}$ minimizations the 50 smallest and
  the 50 largest values for $\Omega_{m}$ and $\Omega_{\Lambda}$. The
  simulated measurement uncertainties of the Einstein angle (\emph{on
  the top}) and of the central velocity dispersion (\emph{on the
  bottom}) range from 4\% to 7\% (a nominal $0\%$ uncertainty is
  assigned to one of the two quantities and the error on the other is
  varied) for samples of 100 (\emph{on the left}) and 200 (\emph{on
  the right}) lenses.
\end{list}
\end{table*}

\begin{table*}
\caption{Intervals at 95\% confidence level for the matter and the
  dark-energy density parameters in a flat cosmology
  ($\Omega_{m}+\Omega_{\Lambda}=1$).}
\label{table:2}
\centering
\begin{tabular}{c c c c c c c c c c}
\hline\hline & & N=100 & & & & & N=200 & & \\ \hline
Err. $\theta_{\mathrm{E}}$ & $4\%$ & $5\%$ & $6\%$ & $7\%$ &
Err. $\theta_{\mathrm{E}}$ & $4\%$ & $5\%$ & $6\%$ & $7\%$ \\ \hline
$\Omega_{m}$ & $\left[0.28,0.34\right]$ & $\left[0.28,0.36\right]$ &
$\left[0.28,0.38\right]$ & $\left[0.28,0.40\right]$ & $\Omega_{m}$ &
$\left[0.29,0.33\right]$ & $\left[0.29,0.34\right]$ &
$\left[0.29,0.36\right]$ & $\left[0.30,0.38\right]$ \\
$\Omega_{\Lambda}$ & $\left[0.66,0.72\right]$ &
$\left[0.64,0.72\right]$ & $\left[0.62,0.72\right]$ &
$\left[0.60,0.72\right]$ & $\Omega_{\Lambda}$ &
$\left[0.67,0.71\right]$ & $\left[0.66,0.71\right]$ &
$\left[0.64,0.71\right]$ & $\left[0.62,0.70\right]$ \\ \hline 
Err. $\sigma_{0}$ & $4\%$ & $5\%$ & $6\%$ & $7\%$ &
Err. $\sigma_{0}$ & $4\%$ & $5\%$ & $6\%$ & $7\%$ \\ \hline
$\Omega_{m}$ & $\left[0.26,0.38\right]$ & $\left[0.26,0.41\right]$ &
$\left[0.26,0.44\right]$ & $\left[0.26,0.49\right]$ & $\Omega_{m}$ &
$\left[0.28,0.36\right]$ & $\left[0.28,0.39\right]$ & 
$\left[0.28,0.41\right]$ & $\left[0.29,0.45\right]$ \\
$\Omega_{\Lambda}$ & $\left[0.62,0.74\right]$ &
$\left[0.59,0.74\right]$ & $\left[0.56,0.74\right]$ &
$\left[0.51,0.74\right]$ & $\Omega_{\Lambda}$ & $\left[0.64,0.72\right]$ &
$\left[0.61,0.72\right]$ & $\left[0.59,0.72\right]$ &
$\left[0.55,0.71\right]$ \\ \hline 
\end{tabular}
\begin{list}{}{}
\item[Notes --]These intervals are obtained by
  excluding from the 2000 $\chi^{2}$ minimizations the 50 smallest and
  the 50 largest values for $\Omega_{m}$ and $\Omega_{\Lambda}$. The
  simulated measurement uncertainties of the Einstein angle (\emph{on
  the top}) and of the central velocity dispersion (\emph{on the
  bottom}) range from 4\% to 7\% (a nominal $0\%$ uncertainty is
  assigned to one of the two quantities and the error on the other is
  varied) for samples of 100 (\emph{on the left}) and 200 (\emph{on
  the right}) lenses.
\end{list}
\end{table*}

In order to explore the precision with which the above technique can
probe the cosmological parameters, we have performed several
simulations. We modelled each lens as a singular isothermal sphere
with an external shear component. Because of the known degeneracy
between external shear and ellipticity (Witt \& Mao \cite{witt}), we
remark that our modeling choice is also good at describing simulations
with singular isothermal ellipsoid models. As suggested by real
lensing systems, we considered the lens redshift uniformly distributed
between 0 and 1 and the source redshift between 1 and 3.5; the lens
velocity dispersion and the external shear values were drawn from
uniform distributions ranging from 100 to 350 $\mathrm{km\,s}^{-1}$
for the first, from 0 to 0.2 in magnitude and from 0$\degr$ to
180$\degr$ in orientation for the second (the role of the external
shear component will become relevant in Sect. 4.1). We calculated the
Einstein angles from Eq. (\ref{eq:1}) in a
$(\Omega_{m},\Omega_{\Lambda}) = (0.3,0.7)$ cosmology and completed
each lensing system simulating randomly, inside a square of side
$\theta_{\mathrm{E}}$ and centered on the lens, the position of a
source. Only the lenses with two images of this source and with
$\theta_{\mathrm{E}}$ greater than $0.5''$ were accepted for the next 
analyses. The reason is that we wanted to investigate multiple image
systems similar to those observed, where the images are far enough
from the center of the galaxy acting as a lens to be resolved with the
present technology. 

We started by examining the dependence of the error estimates on
$\Omega_{m}$ and $\Omega_{\Lambda}$ on the simulated uncertainties
on the Einstein angle and on the central velocity dispersion. In order
to do this, we minimized a chi-square ($\chi^{2}$) like estimator
with respect to the two cosmological parameters. This function is
defined by comparing the ``observational'' distance ratio calculated
from $\theta_{\mathrm{E}}$ and $\sigma_{0}$, that is the quantity on
the left in Eq. (\ref{eq:2}), and the ``theoretical'' distance ratio
obtained from the lens and source redshifts, that is the quantity on
the right in Eq. (\ref{eq:2}):
\begin{equation}
\chi^{2}(\Omega_{m},\Omega_{\Lambda})=\sum_{i=1}^{N}\frac{\bigg(\frac{c^{2}}{4\pi}\,\frac{\theta_{\mathrm{E}_{i}}}{\sigma_{0_{i}}^{2}}-r(z_{l_{i}},z_{s_{i}};\Omega_{m},\Omega_{\Lambda})\bigg)^{2}}{\Big(\frac{c^{2}}{4\pi}\Big)^{2}\bigg[\Big(\frac{1}{\sigma_{0_{i}}^{2}}\Big)^{2}(\delta\theta_{\mathrm{E}_{i}})^{2}+\Big(\frac{\theta_{\mathrm{E}_{i}}}{\sigma_{0_{i}}^{4}}\Big)^{2}(\delta\sigma_{0_{i}}^{2})^{2}\bigg]}\,.
\label{eq:3}
\end{equation}
We included realistic measurement errors on the Einstein angle
($\delta\theta_{\mathrm{E}}$) and on the central velocity dispersion
($\delta\sigma_{0}$), considering normal distributions with standard 
deviations equal to $4\%$, $5\%$, $6\%$, and $7\%$ of the true
values. At first, we performed 2000 minimizations for samples of 100
and 200 lenses, assigning a nominal $0\%$ uncertainty to one of the
two quantities and varying the error on the other in the range
reported above. Then, we repeated the analysis just described with the
additional hypothesis of a flat cosmological model
($\Omega_{m}+\Omega_{\Lambda}=1$). Finally, we simulated estimates of
the cosmological parameters, in general and flat cosmology models,
considering 5\% errors on both the Einstein angle and the central
velocity dispersion. This is the precision with which these quantities
can be measured at present by means of the best ground and space-based
telescopes.

The results for the joint and single probability density functions
($f$) of the cosmological parameters are summarized by
Figs. \ref{pa1}, \ref{pa2}, and \ref{pa3} and by Tables \ref{table:1},
\ref{table:2}, and \ref{table:6}. As a first good indication, the
minimum $\chi^{2}$ was found to be asymptotically unbiased,
i.e. centered on the original values $(\Omega_{m},\Omega_{\Lambda}) = 
(0.3,0.7)$ in the limit of small uncertainties on
$\theta_{\mathrm{E}}$ and $\sigma_{0}$. Then, we remark that the true 
values of the cosmological parameters are always inside the regions
and intervals at the 95\% confidence level, but their boundaries are
not symmetrical. In fact, in all the three previously mentioned tables
these confidence intervals for $\Omega_{m}$ are more extended on the
high side; in contrast, the intervals for $\Omega_{\Lambda}$ are
generally larger on the low side. Moreover, since the distance ratio
$r$ depends linearly on the Einstein angle and quadratically on the
central velocity dispersion, larger confidence regions and intervals
are obtained when a fixed uncertainty is considered on the latter
quantity. The assumption about the flatness of the Universe makes our
technique more powerful, allowing a better estimate of the dark-energy
density parameter, as can be seen by comparing Tables \ref{table:1}
and \ref{table:2}. Finally, from the results of Fig. \ref{pa3} and
Table \ref{table:6}, we argue that the current measurement precision
already allows a good estimate of $\Omega_{\Lambda}$ from a
sufficiently large sample of lenses.

\begin{table}
\caption{Intervals at 95\% confidence level for the matter and the
  dark-energy density parameters in general (\emph{second and third
  columns}) and flat (\emph{fourth and fifth columns})
  cosmology.}
\label{table:6}
\centering
\begin{tabular}{ccccc}
\hline\hline & $\Omega_{m}$ & $\Omega_{\Lambda}$ & $\Omega_{m}$ $^{\mathrm{a}}$ &
$\Omega_{\Lambda}$ $^{\mathrm{a}}$ \\ \hline N=100 &
$\left[0.00,1.12\right]$ & $\left[0.52,0.79\right]$ &
$\left[0.27,0.45\right]$ & $\left[0.55,0.73\right]$ \\ N=200 &
$\left[0.00,0.80\right]$ & $\left[0.54,0.74\right]$ & 
$\left[0.29,0.42\right]$ & $\left[0.58,0.71\right]$ \\ \hline
\end{tabular}
\begin{list}{}{}
\item[$^{\mathrm{a}}$] Flat cosmology: $\Omega_{m}+\Omega_{\Lambda}=1$.
\item[Notes --]These intervals are obtained by excluding from the 2000
  $\chi^{2}$ minimizations the 50 smallest and the 50 largest values
  for $\Omega_{m}$ and $\Omega_{\Lambda}$. The simulated measurement
  uncertainties are 5\% of the true values for both the Einstein angle
  and the central velocity dispersion. Samples of 100 and 200 lenses
  are considered.
\end{list}
\end{table}

From what discussed so far, we conclude that this method is especially
well-suited to measure the dark-energy density parameter. This is not
surprising, given the comments made in the previous section about the
different dependence of the distance ratio on the two cosmological
parameters. In view of the fact that, so far, only supernova
observations have been able to reveal directly a significant dark-energy component, our technique offers a promising new test for the
concordance model. It is also particularly interesting to notice that
the 95\% confidence regions on the parameter planes in Fig. \ref{pa1}
and Fig. \ref{pa3} are oriented in such a way to be complementary to
the results of the other cosmological probes currently considered.

\section{Diagnostics of the relevant quantities}

\subsection{The Einstein angle}

\begin{figure}
  \centering
  \includegraphics[width=0.48\textwidth]{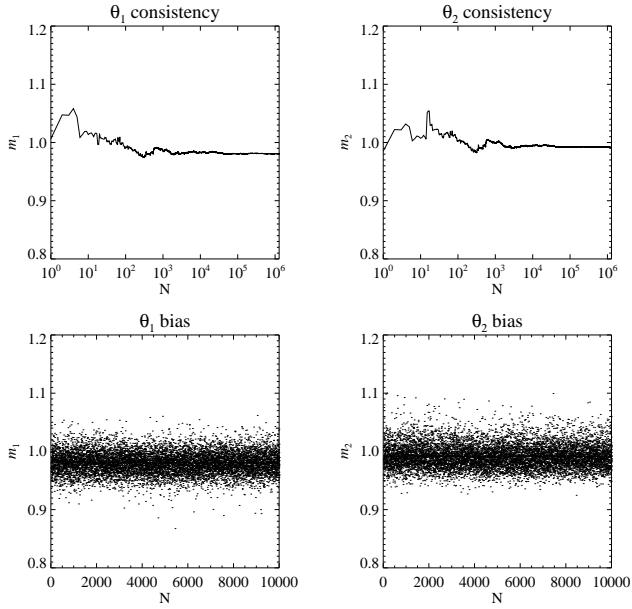}
  \caption{Consistency and bias tests for the $\theta_{\mathrm{E}}$
  estimators. \emph{Top}: Angular coefficients ($m_{1}$ and $m_{2}$)
  of the linear fits of the true Einstein angle versus the estimated
  value from $\theta_{1}$ (\emph{on the left}) and $\theta_{2}$
  (\emph{on the right}), for an increasing number of
  lenses. \emph{Bottom}: Angular coefficients of the linear fits of 
  the true Einstein angle versus the estimated value from $\theta_{1}$
  (\emph{on the left}) and $\theta_{2}$ (\emph{on the right}) for
  10000 samples with 100 lenses each. Solid lines show the mean
  values.}
  \label{es1}%
\end{figure}
\begin{table}
\caption{Statistics of the two estimators of the Einstein angle.} 
\label{table:3}
\centering
\begin{tabular}{ccc}
\hline\hline & Bias & MSE \\ \hline $\theta_{1}$ & $-$0.020 &
0.0008 \\ $\theta_{2}$ & $-$0.008 & 0.0005
\\ \hline
\end{tabular}
\end{table}

Isolated lens galaxies which show low ellipticity values in the
luminous distribution are expected to be well described by
axisymmetric models; on the other hand, non-axisymmetric lens models
are required to represent galaxies in groups or clusters and galaxies
displaying high ellipticity in the luminosity profile. In the former
case, the lens properties are embodied by the so-called Einstein
angle; in the latter case, an equivalent Einstein angle can still be
considered. However, in both cases the uncertainty with which the
Einstein angle can be reconstructed by lensing modeling is on the
order of 5\%, provided high-quality imaging of the systems is
available.

In future cosmological studies we will have to handle data sets
from wide and deep sky surveys made of hundreds of lenses. In order to
evaluate in a simple way $\theta_{\mathrm{E}}$ for lensing systems
with two images of the same source, we may define two estimators. They
could be used to perform a preliminary fast analysis of the data, 
before detailed models with better precision are built. (This problem
has not been taken into account in the measurements of the
cosmological parameters presented in this paper, because in the
following we will consider only data for which refined lens models are
available). The first estimator ($\theta_{1}$) is given by the
semi-distance between the two images, and the second one
($\theta_{2}$) by the semi-sum of the distances of the two images from
the center of the lens. The triangular inequality ensures that
$\theta_{2}$ is always greater than $\theta_{1}$. We remark that the
first quantity is easy to measure, even when the lens (mass) center is
not known a priori; anyway, in most cases the lens center should be
identified with the galaxy luminosity centroid. The two estimators
coincide when a single axisymmetric model is considered, but they
differ when some ellipticity for the lens or an external shear
component is present. In Fig. \ref{es1} and Table \ref{table:3} we
illustrate the statistical properties of $\theta_{1}$ and
$\theta_{2}$, obtained from simulations in which the physical
variables were selected as described in the previous section. The
consistency test and the values of bias and mean squared error (MSE)
(for definitions see Cowan \cite{cowan}) favour $\theta_{2}$ as the
better estimator for $\theta_{\mathrm{E}}$.

\subsection{The velocity dispersion}

The velocity dispersion of stars in galaxies is a well-defined
dynamical quantity (see Bertin \cite{ber3}). The expression
``velocity dispersion'' is also used in lensing studies to refer to a
parameter of the $1/r^{2}$ density distribution characteristic of the
isothermal sphere. In fact, the measured total mass of a lens within
the Einstein angle can be easily converted into an effective velocity
dispersion, given the relation between $\theta_{\mathrm{E}}$ and
$\sigma_{\mathrm{SIS}}$ of Eq.~(\ref{eq:1}). As already mentioned in
Sect. 2, a one-component isothermal model has proved to be an adequate
description for the \emph{total} density distribution in elliptical
galaxies, as far as lensing is concerned. On the other hand, dynamical
modeling requires two-component (luminous+dark) models in order to
find valid agreement with the observations; in principle, the velocity
dispersion of the stellar component might be different from that of
the dark component. Here we wish to compare the central value of the
stellar velocity dispersion to the velocity dispersion of a
one-component isothermal model supposed to represent the total
(stellar and dark) matter distribution. We will find that a good
estimator of the one-component velocity dispersion of an isothermal
model ($\sigma_{\mathrm{SIS}}$) is indeed the stellar central velocity
dispersion ($\sigma_{0}$). This step is essential in order to apply
the method proposed in this paper.

\begin{figure}
  \centering
  \includegraphics[width=0.48\textwidth]{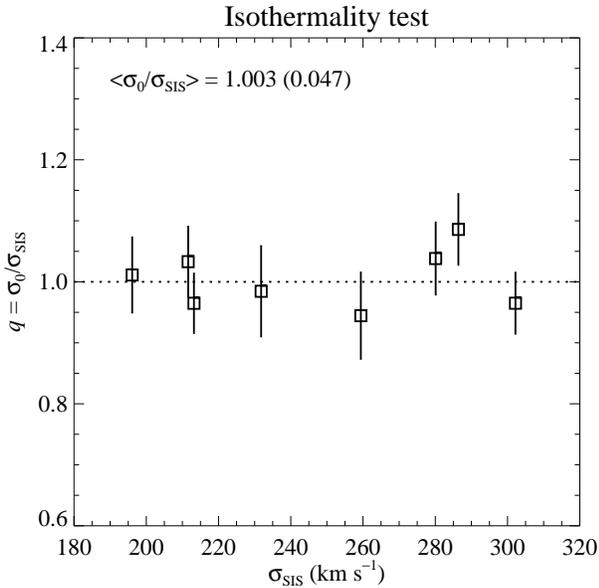}
  \caption{Ratio ($q$) of the stellar central velocity dispersion
  ($\sigma_{0}$) to the one-component velocity dispersion of the
  singular isothermal sphere ($\sigma_{\mathrm{SIS}}$) that best fits
  the photometry and the kinematics for a sample of eight bright,
  nearly round, early-type galaxies.}
  \label{dyn}
\end{figure}

We have considered eight bright, nearly round, early-type galaxies,
which were modeled as described in Bertin et al. (\cite{ber1}) and
Saglia et al. (\cite{sag}). We assumed that the total mass of each
galaxy could be described in terms of a singular isothermal sphere and
calculated $\sigma_{\mathrm{SIS}}$ starting from the best-fit
two-component models of Saglia et al. (\cite{sag}). Then, using the
kinematical data reported in Davies \& Birkinshaw (\cite{dav}) and
Franx et al. (\cite{fra}), we estimated $\sigma_{0}$  for the same
galaxy sample. The results are summarized in Table \ref{table:4}; the
ratio of the two velocities ($q=\sigma_{0}/\sigma_{\mathrm{SIS}}$) is
displayed in Fig. \ref{dyn}. The uncertainty on $\sigma_{0}$ is just
the rms scatter of the velocity dispersions measured at different slit
position angles, so it is probably an underestimate of the real
value. In addition to this, a minimum 5\% error on the
$\sigma_{\mathrm{SIS}}$ was assumed in order to obtain the error bars
of Fig. \ref{dyn}. We notice that $q$ has an average value very close
to unity ($1.003 \pm 0.017$) and an intrinsic rms scatter remarkably
small (0.047). Therefore, we infer that $\sigma_{0}$, which is related 
to the stellar component alone, is a good diagnostic of
$\sigma_{\mathrm{SIS}}$, a tracer of the total (stellar and dark)
matter distribution. Moreover, the ratio $q$ does not show any
significant dependence either on the $\sigma_{\mathrm{SIS}}$ value, as
can be seen in Fig. \ref{dyn}, or on the dark matter fraction
$f_{\mathrm{DM}}$ inside $R_{\mathrm{e}}$, as reported in Table
\ref{table:4}. This fact suggests the existence of an efficient
mechanism of coupling between stellar and dark mass
(``conspiracy''). In addition, we also remark that $q$ is not
influenced by a particular choice of the cosmological parameter
values, because the galaxy set is placed in the nearby Universe.

\begin{table}
\caption{Velocity dispersions and dark matter fraction for a sample of eight bright, nearly round,
  early-type galaxies. }
\label{table:4}
\centering
\begin{tabular}{ccccc}
\hline\hline Object &NGC 1404&NGC 1549&NGC 3379&NGC 4278\\\hline
$\sigma_{0}$ &  245 $\pm$ 7 & 198 $\pm$ 6 & 219 $\pm$ 13 & 228 $\pm$ 9
\\ $\sigma_{\mathrm{SIS}}$ & 259 & 196 & 212 & 232 \\ $f_{\mathrm{DM}}$
& 0.58 & 0.59 & 0.40 & 0.18 \\ \noalign{\smallskip} \hline\hline
Object &NGC 4374&NGC 4472&NGC 4486&NGC 4636\\ \hline 
$\sigma_{0}$ & 291 $\pm$ 6 & 292 $\pm$ 7 & 311 $\pm$ 3 & 206 $\pm$ 8
\\ $\sigma_{\mathrm{SIS}}$ & 280 & 302 & 286 & 213 \\ $f_{\mathrm{DM}}$
& 0.59 & 0.33 & 0.34 & 0.27 \\\hline
\end{tabular}
\begin{list}{}{}
\item[Notes --]Stellar central velocity dispersion ($\sigma_{0}$),
  one-component velocity dispersion of the singular isothermal sphere
  ($\sigma_{\mathrm{SIS}}$) that best fits the photometry and the
  kinematics, and dark matter mass fraction inside $R_{\mathrm{e}}$
  ($f_{\mathrm{DM}}$).
\item[References --]Davies \& Birkinshaw (\cite{dav}); Franx et
  al. (\cite{fra}); Saglia et al. (\cite{sag}).
\end{list}
\end{table}

A larger sample (the SLACS sample, described in the following section)
of more distant early-type galaxies, but with essentially the same
range of velocity dispersion, was studied similarly by Treu et
al. (\cite{tre1}). There they estimated the isothermal velocity 
dispersion inside the Einstein radius $R_{\mathrm{E}}$ from the best
lensing model and their results about the value and intrinsic
scatter of $q$ are consistent with what we found. In conclusion, since
we demonstrated that $\sigma_{0}$ is a good estimator of
$\sigma_{\mathrm{SIS}}$, we argue that the former quantity can be
substituted in Eq. (\ref{eq:1}) to get the cosmology-dependent
relation given by Eq. (\ref{eq:2}).

\section{The SLACS and LSD samples}

\begin{table}
\caption{The lens samples of the SLACS (1-15) and LSD (16-20) Surveys.}
\label{table:5}
\centering
\begin{tabular}{cccccc}
\hline\hline \# & $z_{l}$ & $z_{s}$ & $\theta_{\mathrm{E}}$ (\arcsec)
& $\sigma_{0}$ (km s$^{-1}$) & $R_{\mathrm{e}}$ (\arcsec) \\ \hline 1 & 0.1955 & 0.6322 & 1.47 &
282 $\pm$ 11 & 2.38 $\pm$ 0.02 \\ 2 & 0.3317 & 0.5235 & 1.15 & 349 $\pm$ 24 & 3.37 $\pm$ 0.22 \\ 3 &
0.3223 & 0.5812 & 1.03 & 326 $\pm$ 16 & 3.26 $\pm$ 0.13 \\ 4 & 0.1642 & 0.3240 & 1.61 &
325 $\pm$ 12 & 4.81 $\pm$ 0.02 \\ 5 & 0.2405 & 0.4700 & 1.32 & 318 $\pm$ 17 & 2.60 $\pm$ 0.03 \\ 6 &
0.1260 & 0.5349 & 1.00 & 229 $\pm$ 13 & 1.82 $\pm$ 0.05 \\ 7 & 0.2318 & 0.7950 & 1.15 &
274 $\pm$ 15 & 1.77 $\pm$ 0.01 \\ 8 & 0.0808 & 0.7115 & 0.85 & 195 $\pm$ 10 & 1.23 $\pm$ 0.01 \\ 9 &
0.2046 & 0.4814 & 1.39 & 290 $\pm$ 16 & 3.14 $\pm$ 0.02 \\ 10 & 0.0629 & 0.5352 & 1.04 &
206 $\pm$  5 & 2.60 $\pm$ 0.10\\ 11 & 0.2076 & 0.5241 & 1.21 & 295 $\pm$ 13 & 2.14 $\pm$ 0.02\\ 12 &
0.2479 & 0.7933 & 1.81 & 279 $\pm$ 17 & 2.02 $\pm$ 0.02 \\ 13 & 0.2285 & 0.4635 & 1.25 &
305 $\pm$ 19 & 1.80 $\pm$ 0.01\\ 14 & 0.1553 & 0.5170 & 1.64 & 271 $\pm$ 16 & 4.20 $\pm$ 0.04\\ 15 &
0.0819 & 0.5324 & 1.57 & 245 $\pm$  7 & 4.47 $\pm$ 0.01 \\ \hline \noalign{\smallskip}
16 & 0.485 & 3.595 & 1.34 & 229 $\pm$ 15 & 0.82 $\pm$ 0.12 \\ 17 & 0.938 & 2.941 
& 1.24 & 251 $\pm$ 19 & 1.60 $\pm$ 0.15 \\ 18 & 0.810 & 3.399 & 1.41 & 224 $\pm$ 15 & 1.06 $\pm$ 0.08 \\ 19
& 0.497 & 2.092	& 0.36 & 116 $\pm$ 10 & 0.41 $\pm$ 0.04 \\ 20 & 1.004 & 3.263 & 1.56 & 328
$\pm$ 32 & 0.31 $\pm$ 0.06 \\ \hline
\end{tabular}
\begin{list}{}{}
\item[References --]Treu et al. (\cite{tre3}), (\cite{tre4}), (\cite{tre2}),
  (\cite{tre1}); Koopmans et al. (\cite{koo1}), (\cite{koo2}), (\cite{koo3}).
\end{list}
\end{table}

\begin{figure}
  \centering
  \includegraphics[width=0.24\textwidth]{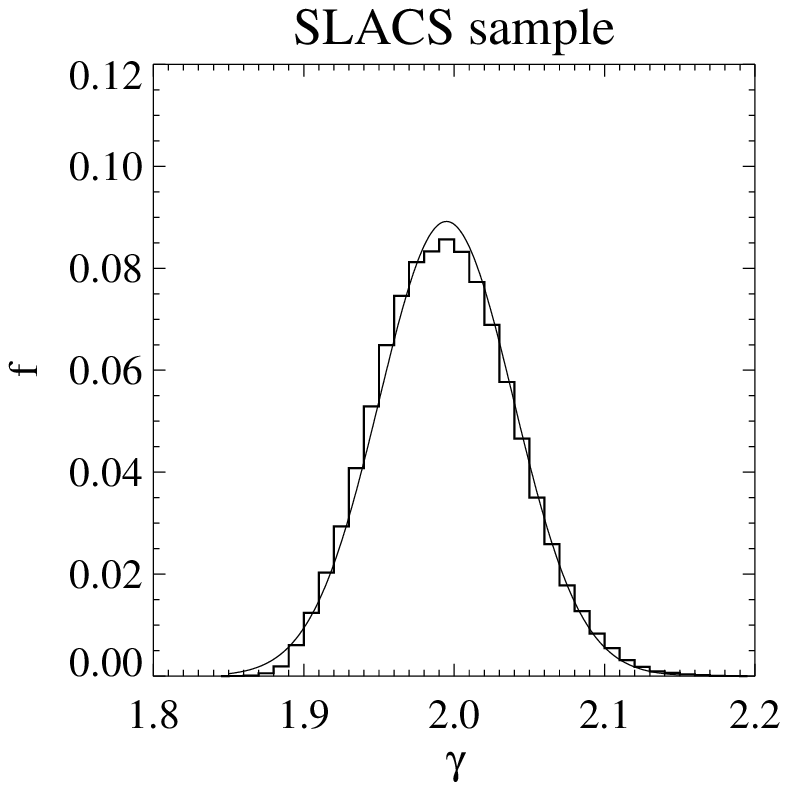}
  \includegraphics[width=0.24\textwidth]{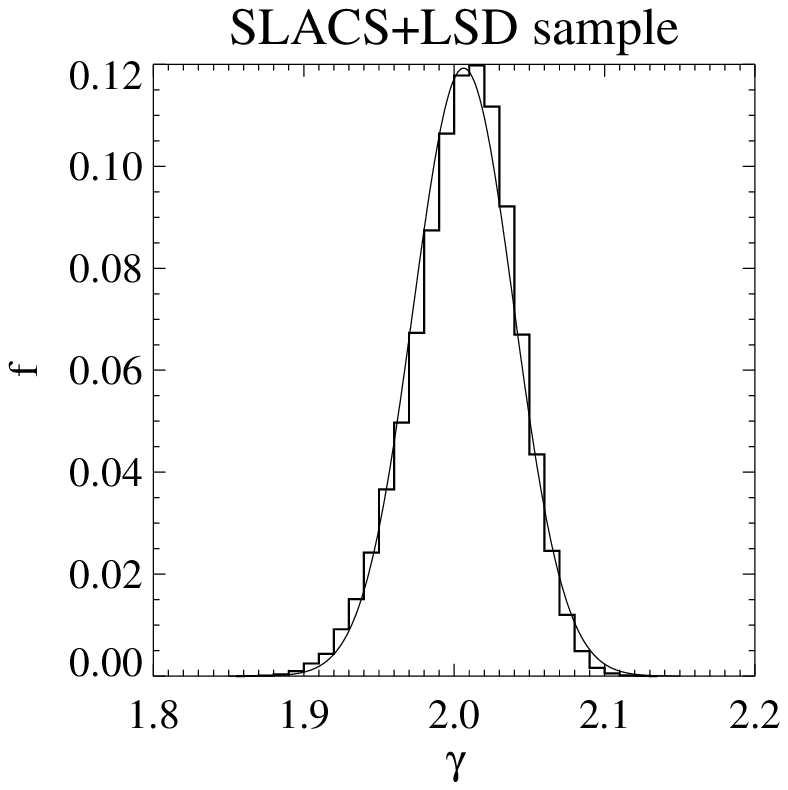}
\caption{Probability density function ($f$), marginalized over the values of the cosmological parameters, of the exponent ($\gamma$) which characterizes the total density profile of the lens galaxies. The probability distribution with the approximate normal distribution for the SLACS (\emph{on the left}) and for the SLACS+LSD (\emph{on the right}) samples are obtained through MCMC methods.}
 \label{fig:7}
\end{figure}

As an application of our technique, we considered the SLACS
(\emph{Sloan Lens ACS}) Survey data set presented in Bolton et
al. (\cite{bol}), Treu et al. (\cite{tre1}), and Koopmans et
al. (\cite{koo3}) (see Table~\ref{table:5}). The sample is composed of
fifteen massive, early-type, lens galaxies at redshifts $z_{l}$
ranging from 0.06 to 0.33 and background sources with a redshift range
of $z_{s}$ from 0.32 to 0.80. Detailed gravitational lensing models on
the \emph{HST/ACS} images were developed in order to measure the total
mass inside the Einstein angle, i.e. $\theta_{\mathrm{E}}$ for an
isothermal model, to a less than a few percent accuracy; from the same images also the values of the galaxy effective radii ($R_{\mathrm{e}}$) were derived. Furthermore,
$\sigma_{0}$ and its uncertainty (average value of 5.0\%) were
obtained from the \emph{Sloan Digital Sky Survey Database}. Afterwards, we studied a second data set, adding to the previous one five additional field elliptical galaxies from the LSD (\emph{Lenses
  Structure and Dynamics}) Survey (Koopmans \& Treu \cite{koo1},
\cite{koo2}; Treu \& Koopmans \cite{tre3}, \cite{tre4}, \cite{tre2})
(see Table \ref{table:5}). These lens and source galaxies are at
higher redshifts ($z_{l}\approx0.5-1.0$, $z_{s}\approx2.0-3.6$), thus
it follows from Sect. 2 that the dependence of $r$ on the cosmological
parameters should be easier to test. On the other hand, these galaxies
have higher central velocity dispersion errors, so that the mean
uncertainty on $\sigma_{0}$ rises now to 5.7\%. 

At first, we checked on the two samples the plausibility of the hypothesis of homologous total density distribution, without any assumptions on the values of the cosmological parameters. In order to do so, we took into account the following relation (Koopmans \cite{koo5})
\begin{equation}
\frac{c^{2}}{4\pi}\,\frac{\theta_{\mathrm{E}}}{\sigma_{0}^{2}}=r(z_{l},z_{s};\Omega_{m},\Omega_{\Lambda})\bigg(\frac{8\theta_{\mathrm{E}}}{R_{\mathrm{e}}}\bigg)^{2-\gamma}\frac{\gamma(3-\gamma)}{2}\,,
\label{eq:4}
\end{equation}
which extends Eq. (\ref{eq:2}), considering a more general power law model for the total density profile ($\rho \propto 1/r^{\gamma}$) of the lens galaxies (the isothermal case is retrieved setting $\gamma$ equal to 2). The parameter $\gamma$ was assumed to be the same for all the lenses and we used Markov chain Monte Carlo (MCMC) methods (with $5 \times 10^{5}$ steps for each chain) to sample its probability density distribution. The marginalized probability density function ($f$) with the approximate normal distribution, the mean value, and the standard deviation of $\gamma$ for the two SLACS and SLACS+LSD samples are shown in Fig. \ref{fig:7} and Table \ref{table:7}. From these results, we can state that the total density profile of the lens ellipticals is indeed well approximated by an isothermal distribution ($\gamma$ equal to 2), independentely on the adopted cosmological model.

\begin{table}
\caption{Mean value and standard deviation of the parameter $\gamma$ for the SLACS and SLACS+LSD samples.} 
\label{table:7}
\centering
\begin{tabular}{ccc}
\hline\hline & SLACS & SLACS+LSD \\ \hline $\gamma$ & 1.99$\pm$0.04 & 2.01$\pm$0.03\\ \hline
\end{tabular}
\end{table}

Starting from here, we evaluated the observational angular diameter distance
ratio of Eq. (\ref{eq:2}) and its uncertainty, assuming a reasonable
5\% error on $\theta_{\mathrm{E}}$. Hence, we used the $\chi^{2}$
statistics reported in Eq. (\ref{eq:3}) to gain information about the
cosmological parameters.

\begin{figure}
  \centering
  \includegraphics[width=0.49\textwidth]{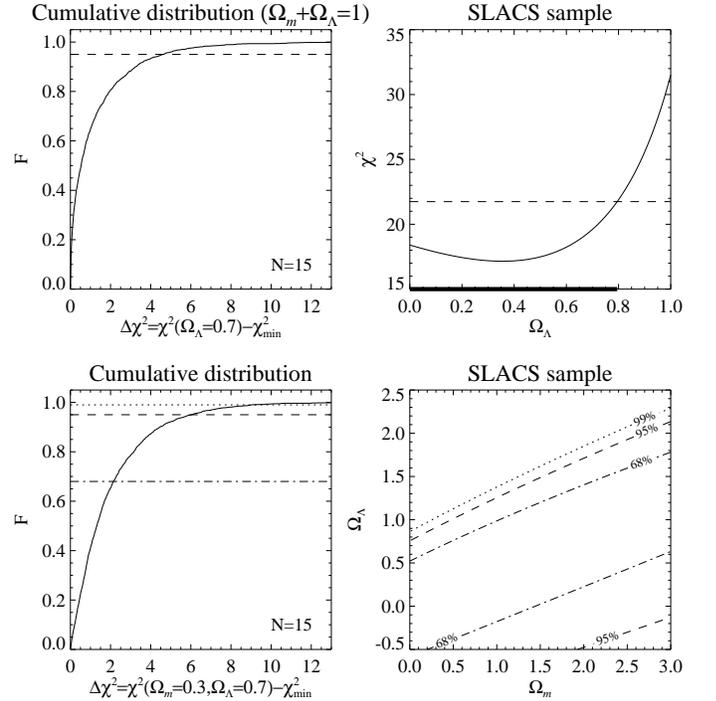}
  \caption{Results from the SLACS sample. \emph{Top left:} Cumulative
  distribution (F) of the distance between the $\chi^{2}$ in the
  true ($\Omega_{\Lambda}=0.7$) and the best flat cosmological model
  from 2000 simulations with 15 lenses each. The dashed line gives the
  95\% confidence level. \emph{Top right:} The $\chi^{2}$ curve for
  the SLACS sample in a flat cosmology. The dashed line corresponds to
  the value of $\Delta\chi^{2}$ of the previous panel and the thick
  bar on the abscissa axis shows the 95\% confidence level interval
  for $\Omega_{\Lambda}$. \emph{Bottom left:} Cumulative distribution
  (F) of the distance between the $\chi^{2}$ in the true
  $(\Omega_{m},\Omega_{\Lambda})=(0.3,0.7)$ and the best cosmological 
  model from 2000 simulations with 15 lenses each. The dashed-dotted,
  dashed, and dotted lines give, respectively, the 68\%, 95\%, and
  99\% confidence levels. \emph{Bottom right:} The $\chi^{2}$ contour
  plot for the SLACS sample; the levels correspond to the values of
  $\Delta\chi^{2}$ of the previous panel. The dashed-dotted, dashed,
  and dotted lines show, respectively, the 68\%, 95\%, and 99\%
  confidence regions for $\Omega_{m}$ and $\Omega_{\Lambda}$.}
  \label{SL1}
\end{figure}

\begin{figure}
  \centering
  \includegraphics[width=0.49\textwidth]{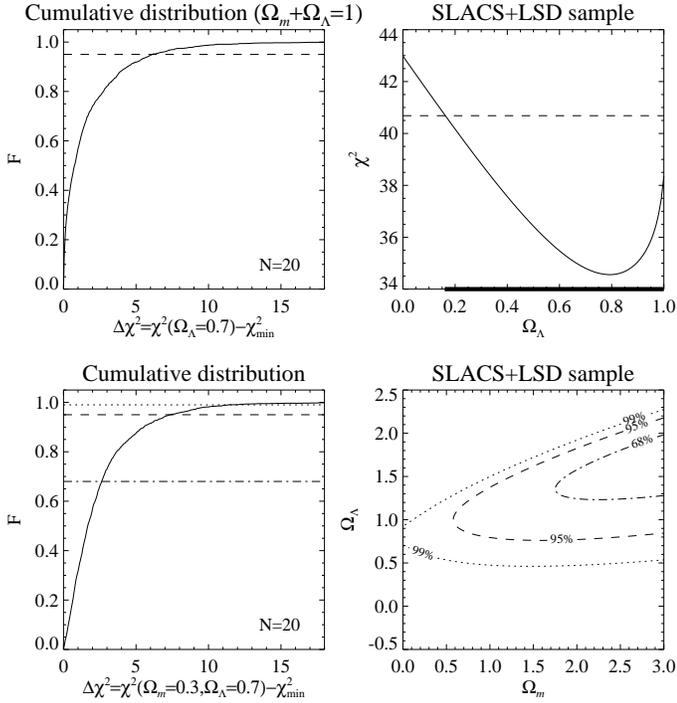}
  \caption{Results from the SLACS+LSD sample. \emph{Top left:}
  Cumulative distribution (F) of the distance between the $\chi^{2}$
  in the true ($\Omega_{\Lambda}=0.7$) and the best flat cosmological
  model from 2000 simulations with 20 lenses each. The dashed line
  gives the 95\% confidence level. \emph{Top right:} The $\chi^{2}$
  curve for the SLACS+LSD sample in a flat cosmology. The dashed line
  corresponds to the value of $\Delta\chi^{2}$ of the previous panel
  and the thick bar shows the 95\% confidence level interval for
  $\Omega_{\Lambda}$. \emph{Bottom left:} Cumulative distribution
  (F) of the distance between the $\chi^{2}$ in the true
  $(\Omega_{m},\Omega_{\Lambda})=(0.3,0.7)$ and the best cosmological
  model from 2000 simulations with 20 lenses each. The dashed-dotted,
  dashed, and dotted lines give, respectively, the 68\%, 95\%, and
  99\% confidence levels. \emph{Bottom right:} The $\chi^{2}$ contour
  plot for the SLACS+LSD sample; the levels correspond to the values
  of $\Delta\chi^{2}$ of the previous panel. The dashed-dotted,
  dashed, and dotted lines show, respectively, the 68\%, 95\%, and
  99\% confidence regions for $\Omega_{m}$ and $\Omega_{\Lambda}$.}
  \label{SL2}
\end{figure}

The results are plotted in Figs. \ref{SL1} and \ref{SL2}. Here we show
the intervals at the 95\% confidence level for the dark-energy density
parameter in a flat cosmological model, and the 68\%, 95\%, and 99\%
confidence regions in the parameter space
($\Omega_{m},\Omega_{\Lambda}$) for general cosmological models. In a
flat geometry, at 95\% CL, $\Omega_{\Lambda}$ is found to be smaller
than 0.80 from the SLACS sample, and greater than 0.16 from the 
SLACS+LSD sample. In addition, without any assumptions on the
cosmological parameters, the SLACS+LSD sample rules out, at greater
than 99\% CL, cosmological models with dark-energy density parameter
smaller than 0.4 (as shown in the bottom right panel of
Fig. \ref{SL2}). From a comparison of Fig. \ref{SL1} and
Fig. \ref{SL2}, it turns out that the larger set of galaxies with more 
distant lenses shifts the minimum $\chi^{2}$ towards higher values of
$\Omega_{\Lambda}$ and that a much bigger region of values for the
cosmological parameters is excluded, at the same confidence level.  

The small number of lenses in the two samples is the primary reason
that prevents us from reaching higher precision results. We note that
the current concordance measurement of $\Omega_{\Lambda}$ suggests a
value between 0.7 and 0.8 and this range is included in the 95\%
confidence intervals and in the 99\% confidence regions for both the
SLACS and the SLACS+LSD samples.

\section{Conclusions}

Lensing studies have been recognized to be very valuable in providing
a testing ground for theories of formation and evolution of early-type
galaxies (Rusin \& Kochanek \cite{rus2}; Koopmans et
al. \cite{koo3}). In these studies the structure of the lens galaxies
can be investigated, once a specific cosmological model is
adopted. From a complementary perspective, a measurement of the
cosmological parameters can in principle be accomplished by assuming
some empirically justified general properties for a sample of
lenses. In this paper, we have shown that indeed, starting from the
$1/r^{2}$ paradigm for the total density profile of elliptical
galaxies, the cosmological parameters $\Omega_{m}$ and
$\Omega_{\Lambda}$ can be estimated. The adopted paradigm has been
validated by stellar dynamics (e.g., Gerhard et al. \cite{ger}),
strong gravitational lensing (e.g., Rusin et al. \cite{rus}), and
dynamical studies based on several other tracers: globular clusters
and planetary nebulae (e.g., Peng et al. \cite{pen}),  X-ray halos
(e.g., Humphrey et al. \cite{hum}), and HI disks and rings (e.g.,
Franx et al. \cite{fra2}).

In detail, the main results of this paper can be summarized as
follows:

\begin{itemize}
\item We have developed a new method to investigate the geometry of
  the Universe by combining measurements of strong gravitational
  lensing and stellar dynamics in a sample of elliptical galaxies. The
  basic idea is to compare lensing and dynamical mass estimates in
  order to find an observable cosmology-dependent relation. For this
  purpose, it is necessary to model the total (luminous and dark)
  density profile inside the Einstein radius of the
  ellipticals under investigation. Here, we have assumed a
  one-component $1/r^{2}$ density profile.

\item We have performed a feasibility study of our technique by
  simulating measurements of the cosmological parameters. We have
  studied the precision of the method, which depends on the level of
  uncertainty with which the Einstein angle and the central stellar
  velocity dispersion are known. Hence, we have demonstrated that the
  current uncertainty with which these quantities can be measured is
  sufficiently small to obtain precise information about cosmology,
  once a statistically significant sample of lenses is available. In
  particular, the method has been shown to be best suited to measure
  $\Omega_{\Lambda}$.

\item Then, we have studied the relation between the stellar
  kinematics and the velocity parameter of the one-component $1/r^{2}$
  (isothermal) model for the total density distribution in
  ellipticals. For a sample of eight bright, round, and nearby
  early-type galaxies, we have shown that the stellar central velocity
  dispersion is a good tracer of the velocity dispersion
  characterizing the one-component model.

\item We have applied the proposed method to the SLACS Survey data
  set. This sample is composed of fifteen massive early-type galaxies 
  at intermediate redshift acting as lenses. In a flat cosmology this
  method leads to a value of $\Omega_{\Lambda}$ lower than 0.80 at
  95\% CL. By including five more distant lenses from the LSD Survey,
  at the same confidence level and in the same flatness hypothesis,
  $\Omega_{\Lambda}$ has been measured to be greater than 0.16. This
  latter sample ruled out, at greater than 99\% CL, general
  cosmological models with values of $\Omega_{\Lambda}$ smaller than
  0.4.
 
\item Finally, we conclude that future surveys expected to identify a
  large number of lenses (one or two hundred) will allow this method
  to measure the values of the cosmological parameters with a
  precision comparable to that of other standard techniques. This will
  be a new and important test for the $\Lambda$\emph{CDM} concordance
  model.
\end{itemize}


\begin{acknowledgements}
This work was partly supported by the Italian MiUR (PRIN 2004). GB and
CG acknowledge the support and hospitality of the Kavli Institute for
Theoretical Physics at the University of California, Santa Barbara.
\end{acknowledgements}


\begin{thebibliography}{}

\bibitem[1994]{bar1} Bartelmann M. \& Weiss A. 1994,
      A\&A, 287, 1

\bibitem[1998]{bar3} Bartelmann M., Huss A., Colberg J., Jenkins A., and Pearce F. 1998,
      A\&A, 330, 1

\bibitem[1999]{bar2} Bartelmann M. \& Schneider P. 1999,
      A\&A, 345, 17

\bibitem[2003]{ben} Bennett C. L., Halpern M., Hinshaw G. et al. 2003,
      ApJS, 148, 1

\bibitem[1992]{ber1} Bertin G., Saglia R. P., and Stiavelli M. 1992,
      ApJ, 384, 423

\bibitem[2000]{ber3} Bertin G. 2000,
      Dynamics of Galaxies
      (Cambridge University Press)

\bibitem[2002]{ber2} Bertin G., Ciotti L., and Del Principe M. 2002,
      A\&A, 386, 149

\bibitem[2006]{bol} Bolton A. S., Burles S., Koopmans L. V. E., Treu T.,
      and Moustakas L. A. 2006,
      ApJ, 638, 703

\bibitem[2003]{bro} Browne I. W. A., Wilkinson P. N., Jackson N. J. F. et al. 2003,
      MNRAS, 341, 13

\bibitem[2001]{bur} Burles S., Nollett K. M., and Turner M. S. 2001,
      ApJ, 552, 1

\bibitem[2005]{col} Cole S., Percival W. J., Peacock J. A. et al. 2005,
      MNRAS, 362, 505

\bibitem[1998]{cowan} Cowan G. 1998,
      Statistical Data Analysis
      (Clarendon Press, Oxford)

\bibitem[2003]{cyb} Cyburt R. H., Ellis J., Fields B. D., and Olive
      K. A. 2003, PhRvD, 67, j3521C

\bibitem[2004]{dal} Dalal N., Holder G., and Hennawi J. F. 2004,
      ApJ, 609, 50

\bibitem[1988]{dav} Davies R. L. \& Birkinshaw M. 1988,
      ApJS, 68, 409

\bibitem[1989]{fra} Franx M., Illingworth G., and Heckman T. 1989,
      ApJ, 344, 613

\bibitem[1994]{fra2} Franx M., van Gorkom J. H., and de Zeeuw T. 1994,
      ApJ, 436, 642

\bibitem[2001]{fre} Freedman W. L., Madore B. F., Gibson B. K. et al. 2001,
      ApJ, 553, 47

\bibitem[2001]{ger} Gerhard O., Kronawitter A., Saglia R. P., and
      Bender R. 2001,
      AJ, 121, 1936

\bibitem[2006]{hum} Humphrey O., Buote D. A., Gastaldello F. et al. 2006,
      ApJ, 646, 899

\bibitem[2003]{jai} Jain B. \& Taylor A. 2003,
      PhRvL, 91, n1302J

\bibitem[1992]{koc3} Kochanek C. S. 1992,
      ApJ, 384, 1

\bibitem[1993]{koc1} Kochanek C. S. 1993,
      ApJ, 419, 12

\bibitem[1994]{koc2} Kochanek C. S. 1994,
      ApJ, 436, 56

\bibitem[1996]{koc4} Kochanek C. S. 1996,
      ApJ, 473, 595

\bibitem[2002]{koo1} Koopmans L. V. E. \& Treu T. 2002,
      ApJ, 568, 5

\bibitem[2003a]{koo2} Koopmans L. V. E. \& Treu T. 2003a,
      ApJ, 583, 606

\bibitem[2003b]{koo4} Koopmans L. V. E., Treu T., Fassnacht C. D.,
      Blandford R. D., and Surpi G. 2003b,
      ApJ, 599, 70

\bibitem[2005]{koo5} Koopmans L. V. E.,
      astro-ph/0511121

\bibitem[2006]{koo3} Koopmans L. V. E., Treu T., Bolton A. S., Burles S.,
      and Moustakas L. A. 2006,
      ApJ, 649, 599

\bibitem[1994]{kor} Kormann R., Schneider P., and Bartelmann M. 1994,
      A\&A, 284, 285

\bibitem[1998]{lin} Link R. \& Pierce M. J. 1998,
      ApJ, 502, 63

\bibitem[1999]{lom} Lombardi M. \& Bertin G. 1999,
      A\&A, 342, 337

\bibitem[2005]{mit} Mitchell J. L., Keeton C. R., Frieman J. A., and
      Sheth R. K. 2005, ApJ, 622, 81

\bibitem[2006]{mor} M\"ortsell E. \& Sunesson C. 2006,
      JCAP, 01, 012

\bibitem[2003]{mye} Myers S. T., Jackson N. J., Browne I. W. A. et al. 2003,
      MNRAS, 341, 1

\bibitem[1997]{myu} Myungshin I., Griffiths R. E., and Ratnatunga
      K. U. 1997, ApJ, 475, 457

\bibitem[2004]{pen} Peng E. W., Ford H. C., and Freeman K. C. 2004,
      ApJ, 602, 705

\bibitem[1999]{per} Perlmutter S., Aldering G., Goldhaber G. et al. 1999,
      ApJ, 517, 565

\bibitem[2003]{ref} Refregier A. 2003,
      ARA\&A, 41, 645

\bibitem[1964]{refs} Refsdal S. 1964,
      MNRAS, 128, 307

\bibitem[1998]{rie1} Riess A. G., Filippenko A. V., Challis P. et al. 1998,
      AJ, 116, 1009

\bibitem[2004]{rie2} Riess A. G., Strolger L.-G., Tonry J. et al. 2004,
      ApJ, 607, 665

\bibitem[2003]{rus} Rusin D., Kochanek C. S., and Keeton C. R. 2003,
      ApJ, 595, 29

\bibitem[2003]{rus2} Rusin D. \& Kochanek C. S. 2005,
      ApJ, 623, 666

\bibitem[1992]{sag} Saglia R. P., Bertin G., and Stiavelli M. 1992,
      ApJ, 384, 433

\bibitem[1992]{schneider} Schneider P., Ehlers J., and Falco
      E. E. 1992, 
      Gravitational Lenses
      (Springer-Verlag, New York)

\bibitem[2004]{sou} Soucail G., Kneib J.-P., and Golse G. 2004,
      A\&A, 417, 33

\bibitem[2003]{spe1} Spergel D. N., Verde L., Peiris H. V. et al. 2003,
      ApJS, 148, 175

\bibitem[2007]{spe2} Spergel D. N., Bean R., Dor\'e O. et al. 2007,
      ApJS, 170, 377

\bibitem[2004]{teg} Tegmark M., Strauss M. A., Blanton M. R. et al. 2004,
      PhRvD, 69, j3501T

\bibitem[2002]{tre5} Treu T., Stiavelli M., Casertano S., M\o ller P., and Bertin G. 2002,
      ApJ, 564, 13

\bibitem[2002]{tre3} Treu T. \& Koopmans L. V. E. 2002,
      ApJ, 575, 87

\bibitem[2003]{tre4} Treu T. \& Koopmans L. V. E. 2003,
      MNRAS, 343, 29

\bibitem[2004]{tre2} Treu T. \& Koopmans L. V. E. 2004,
      ApJ, 611, 739

\bibitem[2006]{tre1} Treu T., Koopmans L. V. E., Bolton A. S., Burles S.,
      and Moustakas L. A. 2006,
      ApJ, 640, 662

\bibitem[1997]{witt} Witt H. J. \& Mao S. 1997,
      MNRAS, 291, 211

\end{thebibliography}
\end{document}